\documentclass[
]{ceurart}

\usepackage{listings}
\usepackage{graphicx}
\usepackage{textcomp}
\usepackage{xcolor}
\usepackage{subcaption}

\usepackage{mathtools}

\usepackage{multirow}
\usepackage{rotating}
\usepackage{float}
\usepackage{xspace}
\usepackage{nicematrix}
\usepackage{makecell}
\usepackage{nicefrac}
\usepackage{enumitem}
\usepackage{booktabs}
\usepackage{microtype}      % microtypography
\usepackage{amsmath}
\usepackage{amssymb}
\usepackage{amsfonts}
\usepackage{tikz}
\usepackage{pgfplots}
\usepackage{amsthm}
\usepackage{multirow}
\usepackage{colortbl} 
\usepackage{arydshln}
\usepackage{siunitx}
\usepackage{cleveref}
\usepackage{arydshln}
\usepackage{amssymb}
\usepackage{comment}
\usepackage{natbib}  % DO NOT CHANGE THIS AND DO NOT ADD ANY OPTIONS TO IT

\newcommand{\vct}[1]{\ensuremath{\boldsymbol{\mathbf{#1}}}}

\newcommand{\set}[1]{\ensuremath{\mathcal{#1}}}

\DeclareMathOperator*{\minimize}{minimize}

\DeclareMathOperator*{\subjectto}{subject~to}

\newcommand{\myparagraph}[1]{\noindent\textbf{#1}}
\makeatletter
\DeclareRobustCommand\onedot{\futurelet\@let@token\@onedot}
\def\@onedot{\ifx\@let@token.\else.\null\fi\xspace}

\def\eg{\emph{e.g}\onedot}

\makeatother

\newcommand{\ddn}{DDN\xspace}

\newcommand{\apgd}{APGD\xspace}

\newcommand{\pdpgd}{PDPGD\xspace}
\newcommand{\vfga}{VFGA\xspace}

\newcommand{\sigmazero}{\ensuremath{\sigma}-zero\xspace}

\newcommand{\ab}{AttackBench\xspace}

\newcommand{\pertsize}{\ensuremath{\varepsilon}\xspace}

\newcommand{\asr}{\ensuremath{\text{ASR}}}%_\pertsize}}

\theoremstyle{definition}
\begin{document}

%%
%% Rights management information.
%% CC-BY is default license.
\copyrightyear{2025}
\copyrightclause{Copyright for this paper by its authors.
Use permitted under Creative Commons License Attribution 4.0
  International (CC BY 4.0).}

%%
%% This command is for the conference information
\conference{Ital-IA 2025: 5th National Conference on Artificial Intelligence, organized by CINI, June 23-24, 2025, Trieste, Italy}

%%
%% The "title" command
\title{Evaluating the Evaluators: Trust in Adversarial Robustness Tests}%Empirical and Formal Adversarial Robust ML Evaluation Under the AI Act}

%\tnotemark[1]
%\tnotetext[1]{You can use this document as the template for preparing your
%  publication. We recommend using the latest version of the ceurart style.}

%%
%% The "author" command and its associated commands are used to define
%% the authors and their affiliations.
\author[1]{Antonio Emanuele Cinà}[%
orcid=0000-0003-3807-6417,
email=antonio.cina@unige.it,
url=https://cinofix.github.io,
]
\fnmark[1]
\cormark[1]
\address[1]{DIBRIS - Department of Informatics, Bioengineering, Robotics and Systems Engineering, University of Genoa}
\address[2]{Department of Environmental Sciences, Informatics and Statistics, Ca' Foscari University of Venice}

\author[2]{Maura Pintor}[%
orcid=0000-0003-3287-7352,
email=maura.pintor@unica.it,
url=https://maurapintor.github.io,
]

\author[1]{Luca Demetrio}[%
orcid=0000-0001-5104-1476,
email=luca.demetrio@unige.it,
url=https://zangobot.github.io,
]

\author[2]{Ambra Demontis}[%
orcid=0000-0001-9318-6913,
email=ambra.demontis@unica.it,
url=https://web.unica.it/unica/page/it/ambra\_demontis,
]

\author[2]{Battista Biggio}[%
orcid=0000-0001-7752-509X,
email=antonio.cina@unige.it,
url=https://battistabiggio.github.io,
]

\author[1]{Fabio Roli}[%
orcid=0000-0003-4103-9190,
email=fabio.roli@unige.it,
url=https://www.saiferlab.ai/people/fabioroli,
]
%\fnmark[1]

%% Footnotes
\cortext[1]{Corresponding author.}
\fntext[1]{These authors contributed equally.}

%%
%% The abstract is a short summary of the work to be presented in the
%% article.
\begin{abstract}
Despite significant progress in designing powerful adversarial evasion attacks for robustness verification, the evaluation of these methods often remains inconsistent and unreliable. Many assessments rely on mismatched models, unverified implementations, and uneven computational budgets, which can lead to biased results and a false sense of security. Consequently, robustness claims built on such flawed testing protocols may be misleading and give a false sense of security.
As a concrete step toward improving evaluation reliability, we present AttackBench, a benchmark framework developed to assess the effectiveness of gradient-based attacks under standardized and reproducible conditions. AttackBench serves as an evaluation tool that ranks existing attack implementations based on a novel \emph{optimality} metric, which enables researchers and practitioners to identify the most reliable and effective attack for use in subsequent robustness evaluations.
The framework enforces consistent testing conditions and enables continuous updates, making it a reliable foundation for robustness verification. 
\end{abstract}

%%
%% Keywords. The author(s) should pick words that accurately describe
%% the work being presented. Separate the keywords with commas.

%%
%% This command processes the author and affiliation and title
%% information and builds the first part of the formatted document.
\maketitle

\section{Introduction}
In recent years, the growing importance of adversarial robustness has led to the development of numerous evasion attacks~\cite{biggio13-ecml,szegedy14-iclr-intriguing} aimed at crafting adversarial examples with increasing precision and efficiency~\cite{carlini17-sp, chen2018ead, rony2019decoupling, pintor2021fast, Cina2024SigmaZero,zheng2023hardening}. 
These attacks are essential tools to assess how well a model can resist against worst-case perturbations from external malicious users.  As a result, they have become central to evaluating the robustness of machine learning systems, particularly in light of emerging regulatory frameworks (e.g., European AI Act~\cite{european_ai_act_2021}), which introduce strict cybersecurity and robustness requirements for high-risk AI systems.\footnote{\url{https://artificialintelligenceact.eu/article/15/}} 
However, while evasion attack algorithms have advanced rapidly, the methods used to evaluate them have not kept pace in terms of rigor or consistency.
Their evaluations often suffer from methodological flaws that undermine their reliability. 
Specifically, we identify three recurring and critical issues: (i) evaluations rely on inconsistent choices of target models and metrics, ranging from fixed-budget success rates~\cite{Croce2019SparseAI} to median perturbation sizes~\cite{brendel2020accurate, pintor2021fast}, which makes cross-paper comparisons unreliable; (ii) attack implementations in public libraries are frequently re-written without validation against the original code, leading to bugs or silent performance degradation~\cite{carlini2019critique,Pintor22IoF}; and (iii) Computational budgets are inconsistently enforced—for example, some attacks exploit internal restarts~\cite{croce2020minimally} or perform additional hyperparameter tuning~\cite{carlini17-sp, chen2018ead}, which gives an unfair advantage to more resource-intensive methods.

Together, these inconsistencies introduce variance that can severely distort robustness assessments, hinder reproducibility, and create a false sense of security. This leads us to a central and urgent question: 
\begin{center}
    \emph{To what extent can we trust the evaluation tests used to certify adversarial robustness?}
\end{center} 
If the tools used to evaluate ML systems are flawed or ineffective, then any robustness guarantees or certification derived from them may be invalid, potentially exposing users to real-world vulnerabilities.
%%%%%%%

As a concrete step toward addressing the unreliability of current robustness evaluations, we present \ab, a benchmark framework developed to systematically assess the effectiveness and efficiency of gradient-based evasion attacks. 
\ab establishes a standardized and impartial evaluation protocol that enables the identification of attack implementations most capable of revealing a model’s true worst-case vulnerabilities under adversarial conditions. 
In this context, reliability refers to an attack’s ability to consistently find adversarial perturbations require minimal distortion to successfully mislead the model while respecting a constrained query budget.
To support this goal, \ab introduces a novel \emph{optimality} metric, which measures how closely each attack approximates the best empirical solution across a diverse set of models and perturbation budgets. 
Lastly, based on this metric, \ab ranks attack implementations according to their effectiveness and efficiency, providing a principled comparison across different threat models. 
The results are published on a continuously updated online leaderboard\footnote{\url{https://attackbench.github.io}}, helping researchers and practitioners select the most reliable and effective attack strategy when evaluating the adversarial robustness of ML models.

\section{Evasion Attacks}\label{sec:gradient-attacks}
Evasion attacks involve manipulating input data at test time to induce misclassification. 
Examples include modifying malware code to evade detection (i.e., to be misclassified as legitimate) and generating adversarial examples in computer vision—images that appear unchanged to humans but deceive deep learning models~\cite{BiggioCMNSLGR13, Carlini017}. 
% questi hanno l'obiettivo di trovare vulnerabilità esistenti interne a modelli già addestrati e sono stati utilizzati per verificare la robustezza alla misclassificazione a test time di un algoritmo. 
Formally, let $\vct x \in [0, 1]^d$ be an input with true label $y \in \{1, ..., C\}$, and let $f(\vct x, \vct \theta)$ denote the prediction of a trained model with parameters $\vct \theta$. 
These attacks typically aim to find a perturbation $\vct \delta$ such that the perturbed input $\vct x' = \vct x + \vct \delta$ leads to misclassification, while remaining within a bounded perturbation norm and valid input space. This objective can be formalized as the following constrained optimization problem:
\begin{align}
    \minimize_{\vct \delta} \quad & \left( L(\vct x + \vct \delta, y; \vct \theta), \, \| \vct \delta \|_p \right) \\
    \subjectto \quad & \vct x + \vct \delta \in [0, 1]^d \, ,
\end{align}
where $L$ is a loss function that penalizes correct classification. 
Popular choices include the negative cross-entropy, the difference of logits~\cite{carlini17-sp}, and the difference of logits ratio~\cite{croce2020reliable}. The perturbation size is typically constrained under $\ell_p$ norms (\eg, $\ell_0$, $\ell_1$, $\ell_2$, $\ell_\infty$), reflecting different adversarial threat models.

This bi-objective formulation reflects a trade-off between misclassification confidence and minimal perturbation. Accordingly, evasion attacks fall into two families: fixed-budget attacks aim to maximize misclassification within a given perturbation bound~\cite{madry18-iclr}, and minimum-norm attacks seek the smallest perturbation that causes misclassification~\cite{szegedy14-iclr-intriguing, brendel2020accurate}.

\subsection{Evaluation Inconsistencies of Robustness}
Despite the vast number of adversarial attacks developed, each claiming improved performance over its predecessors, their evaluation has often lacked standardization across three critical dimensions: (i) the choice of models and evaluation metrics, (ii) the correctness and consistency of attack implementations, and (iii) the fairness of computational budgets.
With respect to the first dimension, attacks are frequently evaluated on different models and datasets using incompatible success criteria—such as the attack success rate at a fixed $\ell_p$ budget~\cite{Croce2019SparseAI} or the median perturbation size~\cite{brendel2020accurate, pintor2021fast}, which hampers meaningful comparisons. For instance, the effectiveness of attacks is commonly measured via the Attack Success Rate (ASR) under a perturbation budget $\epsilon$, formally defined as:
\begin{equation}
     \asr_a(\pertsize)= \frac{1}{|\set D |} \sum_{(\vct x, y) \in \set D} \mathbb I(f(\vct x, \vct \theta)\neq y \quad \wedge \quad \| \vct x_{\text{adv}} - \vct x \|_p \leq \pertsize) \, .
\end{equation}
This metric captures the proportion of input samples in dataset $\set{D}$ for which the attack successfully induces misclassification (i.e., $f(\vct x, \vct \theta)\neq y$) within the allowed norm constraint (i.e., $\| \vct x_{\text{adv}} - \vct x \|_p \leq \pertsize$). However, ASR is highly sensitive to the choice of $\pertsize$; an attack may perform well at one value of $\epsilon$ but poorly at others, limiting the generality of the conclusions drawn.
To overcome the limitations of pointwise evaluation metrics like ASR, robustness evaluation curves~\cite{biggio18wild} are often used (red curve in \autoref{fig:abstract} (2)). These curves show the model's robust accuracy as a function of the perturbation budget $\pertsize$. These curves capture the trade-off between attack strength and the model's resilience over a continuous range of perturbation magnitudes, offering a richer picture of performance than single-point estimates.
A lower area under the robustness evaluation curve means the attack is more effective, as it reduces the model’s accuracy more quickly. However, this metric depends on the model’s starting (clean) accuracy, so it can’t be fairly compared across models with different initial performance.
Concerning the second dimension, many attacks are re-implemented in public libraries without proper validation against the original code, often leading to performance degradation or the introduction of subtle bugs~\cite{carlini2019critique}.

Lastly, regarding the third dimension, attacks differ significantly in their computational demands. 
Some rely on internal restarts~\cite{croce2020minimally}, hyperparameter searches~\cite{carlini17-sp, chen2018ead}, or repeated query evaluations, which can unfairly advantage them in settings without constraints on time or resources.

As a result of all these inconsistencies, researchers and practitioners may unknowingly draw conclusions from flawed comparisons, and thus deploy models with a false sense of security. 
For example, a model certified as robust under suboptimal evaluation attack may still be easily fooled in practice with more advanced attacks, exposing users and stakeholders to unacceptable risks.

\section{The AttackBench Framework}\label{sec:attackbench}
To support the choice of a reliable attack to assess adversarial robustness, we rely on \ab, a benchmark framework specifically designed to test and compare the effectiveness of gradient-based attacks under consistent, fair, and reproducible conditions. Developed in prior work~\cite{cina2024attackbench}, \ab offers a structured and extensible platform to assess whether robustness evaluation methods themselves are reliable, i.e., whether they are close to producing the optimal (i.e., smallest) possible adversarial perturbations within a fixed query budget.
\ab serves as a framework where attacks are evaluated against a common set of models (the \emph{model zoo}) and datasets, using a fixed query budget that counts both forward and backward passes. 
At the core of \ab is the notion of \emph{optimality}. Instead of measuring only whether an attack succeeds at a certain perturbation size $\pertsize$, \ab evaluates how close each attack comes to an empirical best solution, derived by ensembling the results of all tested methods, for $\pertsize$. 
Specifically, for each attack, \ab evaluate their \emph{local optimality} score, which reflects the quality of an attack on a specific model, and the \emph{global optimality} score, which averages this performance across a diverse set of models. 
Subsequently, \ab utilizes these scores to rank attacks, fostering the identification of those that are both reliable and efficient.
Lastly, a key feature of \ab is its ability to support continuous updates, enabling an evolving leaderboard and encouraging ongoing contributions from the research community.

\subsection{AttackBench Internals}\label{subsec:attackbench-modules}
\begin{figure}[htbp]
    \centering
    \includegraphics[width=1\linewidth]{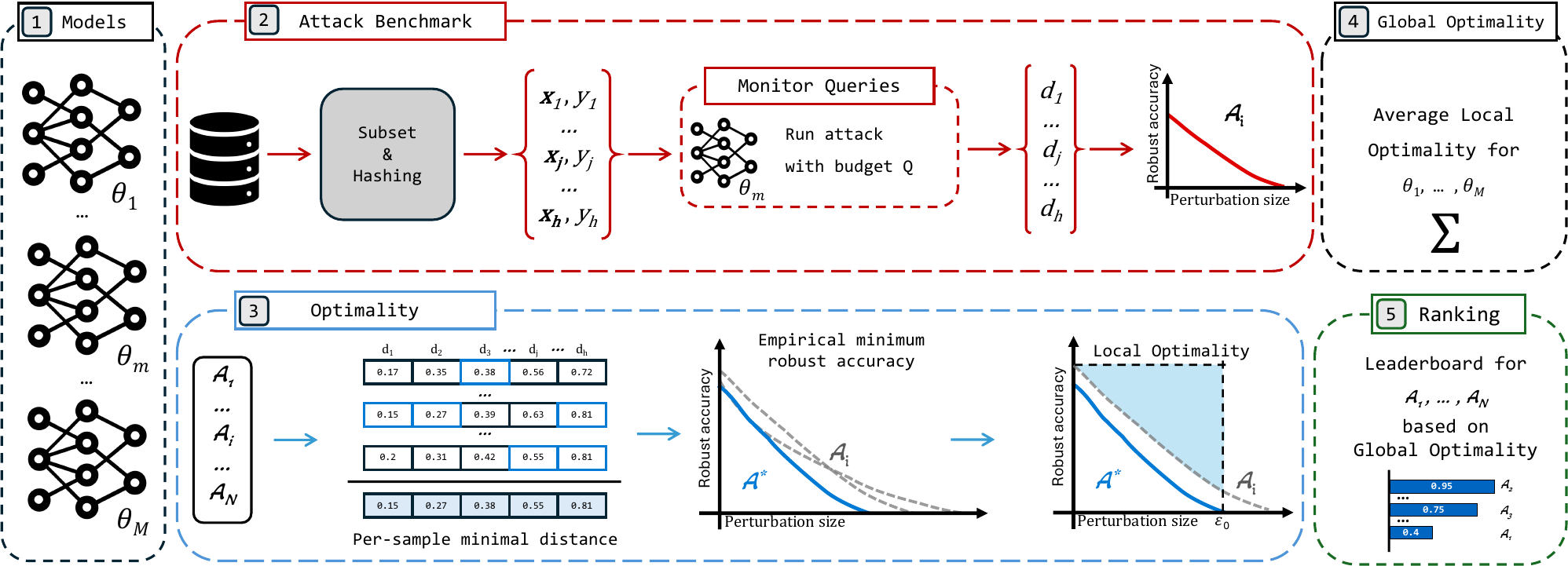}
    \caption{Overview of the five stages of \ab.}
    \label{fig:abstract}
\end{figure}

The framework is organized into five modular stages, each designed to minimize experimental bias and promote reproducibility, depicted in \autoref{fig:abstract}.\medskip

\myparagraph{Stage 1 - Model Zoo.} \ab begins by defining a diverse and extensible model zoo, which includes both robust and standard models. This ensures that attacks are tested across a range of architectures and robustness levels, preventing overfitting to specific models and enabling generalization of benchmarking results.

\myparagraph{Stage 2 - Attack Benchmarking.} Attacks are executed against each model in the zoo under strict constraints, producing, for each model-attack configuration, the corresponding robustness evaluation curve (red curve in \autoref{fig:abstract} (2)). \ab wraps each model in a query-tracking interface that counts both forward and backward passes, ensuring all attacks are evaluated within the same computational budget. Importantly, it records the best adversarial perturbation found within this budget rather than returning the result from the last iteration—an improvement over many existing libraries.\medskip

\myparagraph{Stage 3 – Local Optimality.} To enable meaningful comparisons between different adversarial attacks, \ab introduces the \textit{local optimality} metric—a model-agnostic measure of attack effectiveness. Rather than focusing solely on individual scalar values such as attack success rate at a fixed perturbation size $\pertsize$, this metric evaluates how close an attack comes to the best-known lower bound on robustness, as estimated by aggregating the results of multiple attacks (blue curve in the \autoref{fig:abstract}). Specifically, local optimality is computed from robustness evaluation curves obtained during Stage 2 of each attack.
Specifically, AttackBench ensembles all attacks run against a given model and constructs an empirical lower envelope curve representing the best-known attack performance at each perturbation size. The local optimality score for a specific attack is then calculated as the normalized area under the curve between the attack’s robustness curve and the lower envelope. Formally, the smaller the area between these two curves, the closer the attack is to the best-known bound, and the higher its optimality score. This value is normalized to lie within $[0,1]$, where a score of $1$ indicates that the attack achieves performance indistinguishable from the ensemble lower bound across the full perturbation range. \medskip

\myparagraph{Stage 4 - Global Optimality.} Since local optimality depends on the specific target model, \ab aggregates local scores across all models in the zoo to compute a \textit{global optimality} score. This reflects the average effectiveness of an attack across diverse scenarios, penalizing methods that perform well only on specific architectures. The global score enables ranking attacks in a model-agnostic way.\medskip

\myparagraph{Stage 5 - Ranking and Leaderboard.} Attacks are ranked by their global optimality score and grouped according to the $\ell_p$ threat model they assume. A key advantage of \ab is its incremental update capability: when a new attack is evaluated, only the ensemble statistics and rankings are updated—previous attacks do not need to be re-run. This enables continuous integration and real-time leaderboard updates.

\subsection{Main Take-Home Messages}
\label{sec:main_takeaways}
We now summarize the main take-home messages derived from AttackBench~\cite{cina2024attackbench}. Our benchmarking campaign spans $102$ adversarial attacks, evaluated across $2$ datasets (CIFAR-10 and ImageNet) and $9$ deep neural networks. Lastly, AttackBench offers a comprehensive perspective on attack performance, efficiency, and implementation fidelity across multiple $\ell_p$ threat models.\medskip

\myparagraph{Overall Attack Performance.} Our large-scale evaluation using \ab yields several critical insights into the reliability and practical utility of gradient-based adversarial attacks. First and foremost, our results confirm that a small subset of attacks, i.e., \sigmazero, \ddn, \pdpgd, and \apgd, consistently outperform others across both CIFAR-10 and ImageNet benchmarks. These attacks exhibit high optimality scores and produce robustness evaluation curves that closely track the empirical best attack.\medskip

\myparagraph{Effectiveness-Efficiency Tradeoffs.} Another central observation concerns the effectiveness-efficiency tradeoffs. While high optimality scores are desirable, they do not always imply computational efficiency. For instance, although \apgd demonstrates strong optimality, it incurs higher computational costs compared to \pdpgd, especially on high-dimensional datasets like ImageNet. Conversely, attacks such as \vfga deliver remarkable speed due to early stopping but suffer a notable drop in attack success rate and optimality when scaled to more complex models.\medskip

\myparagraph{Implementation Variability.} Equally important are the discrepancies observed across different implementations of the same attack. Our benchmark reveals significant variations in performance depending on the source library. For example, the \apgd attack implemented in the AdvLib library or its original repository achieves optimal or near-optimal results, whereas the same attack in the ART library shows a drastic performance degradation. Specifically, the optimality drops from $90.9\%$ with the AdvLib implementation to $26\%$ with the ART library on CIFAR-10. We highlight that these inconsistencies are often due to subtle but impactful implementation details, such as the number of restarts or the choice of loss function. These findings underscore the necessity for practitioners to carefully audit attack implementations before using them for model evaluation, as seemingly minor differences can dramatically alter the perceived robustness of a model.\medskip

\myparagraph{Implementation Pitfalls.} Finally, our benchmark identifies several recurring pitfalls in existing libraries. Some attacks crash under specific conditions (e.g., initialization issues, label index bugs), while others fail to support crucial features such as per-sample $\pertsize$ evaluations, compromising the usability of attack tools in practice.\medskip

\section{Conclusion}
In summary, \ab provides a robust and actionable foundation for evaluating the trustworthiness of adversarial attacks. Our findings stress the importance of algorithmic design, implementation rigor, and careful tuning when benchmarking model robustness. They also caution against naive reliance on off-the-shelf attack implementations without thorough validation, especially in safety-critical or regulatory contexts.\medskip

\subsection*{Acknowledgments}
This work has been partially supported by project FISA-2023-00128 funded by the MUR program ``Fondo italiano per le scienze applicate"; the EU—NGEU National Sustainable Mobility Center (CN00000023), Italian Ministry of University and Research Decree n. 1033—17/06/2022 (Spoke 10); the project Sec4AI4Sec, under the EU’s Horizon Europe Research and Innovation Programme (grant agreement no. 101120393); the project ELSA, under the EU’s Horizon Europe Research and Innovation Programme (grant agreement no. 101070617); and projects SERICS (PE00000014) and FAIR (PE0000013) under the MUR NRRP funded by the EU—NGEU. 

\bibliography{sample-ceur}

\end{document}